\documentclass[
]{ceurart}

\usepackage{listings}
\usepackage{colortbl}
\usepackage{amsmath}
\begin{document}

\copyrightyear{2023}
\copyrightclause{Copyright for this paper by its authors. Use permitted under Creative Commons License Attribution 4.0 International (CC BY 4.0)}

\conference{Forum for Information Retrieval Evaluation, December 15-18, 2023, India}

\title{A ML-LLM pairing for better code comment classification}

\tnotetext[1]{Forum for Information Retrieval Evaluation, December 15-18, 2023, India}

\author[1,2]{Hanna Abi Akl}[%
orcid=0000-0001-9829-7401,
email=hanna.abi-akl@dsti.institute]
\address[1]{Data ScienceTech Institute (DSTI),
 4 Rue de la Collégiale 75005 Paris, France}
\address[2]{Université Côte d’Azur, Inria, CNRS, I3S}

\cortext[1]{Corresponding author.}

\begin{abstract}
  The "Information Retrieval in Software Engineering (IRSE) \footnote{https://sites.google.com/view/irse2023/home}" at FIRE 2023 shared task introduces code comment classification, a challenging task that pairs a code snippet with a comment that should be evaluated as either useful or not useful to the understanding of the relevant code. We answer the code comment classification shared task challenge by providing a two-fold evaluation: from an algorithmic perspective, we compare the performance of classical machine learning systems and complement our evaluations from a data-driven perspective by generating additional data with the help of large language model (LLM) prompting to measure the potential increase in performance. Our best model, which took second place in the shared task, is a Neural Network with a Macro-F1 score of 88.401\% on the provided seed data and a 1.5\% overall increase in performance on the data generated by the LLM.
\end{abstract}


\maketitle

\section{Introduction}
In software development, code and documentation go hand-in-hand. Writing code is crucial to maintaining existing code bases, developing new features and fixing bugs. Documentation helps developers make sense of the logic behind written code and provides a steady set of guidelines to iterate over it \cite{10.5555/1388398}. Code commenting is a form of documentation whereby comments written in natural language are inserted in the code \cite{10.5555/1388398}. The advantage of this method is that it helps clarify parts of the code without affecting performance since comments are ignored by compilers \cite{10.5555/1388398}. It also provides an easy way to reflect updates on code changes without having to modify the entire documentation \cite{10.5555/1388398}.

From the practice of writing code comments comes the challenge of identifying useful comments \cite{shinyama2018analyzing}. Writing comments is not always accurate science, and some comments can be outdated or ambiguous \cite{shinyama2018analyzing}. This can be problematic for developers who rely on these comments to understand and alter the code. There is then a real need for code comment checking. This need has framed the task of collecting code comments from real projects in a code-comment database to aid in the task of classifying useful versus not useful comments \cite{majumdar2020comment}. Code comment classification is still a relatively new task that explores the possibility of accurately discriminating between comments that bring added value to the corresponding code and comments that are not pertinent with respect to the surrounding code \cite{10.1145/3574318}.

Recent research has aimed to answer this challenge by compiling a semantic code-comment base by scraping and collecting code and surrounding comments from real projects in C \cite{majumdar2020comment}. Researchers have also explored applying machine and deep learning techniques to solve this binary classification problem by considering useful comments (i.e., informative of the surrounding code) as a class and non-useful comments (i.e., redundant, uninformative or ambiguous) as another \cite{majumdar2022can, ghosh2022information}.

On the other hand, the rise of large language models (LLM) \cite{zhao2023survey} and their ability to pose as a jack-of-all-trades by solving a wide range of machine learning and deep learning problems, coupled with their wealth of training data, make them an interesting entry point for the code comment classification task \cite{kocon2023chatgpt}. Based on the Transformers model \cite{vaswani2023attention}, they are able to create robust embeddings from text, which helps them tackle problems based on natural language \cite{wolf2020huggingfaces}. Another recent breakthrough in LLMs is in generative artificial intelligence, where users combine pre-trained models with different prompting techniques to generate output data (e.g., text) \cite{liu2021pretrain}. This prompting ability is at the heart of prompt engineering, a method that can redirect a LLM into focusing its generation on a specific need. This need can be in the form of answering specific questions, solving certain tasks (e.g., a classification problem) or even producing data in a pre-defined format \cite{zhou2023large}. The latter use case plays a detrimental role in data augmentation, whereby users can couple the power of LLMs with a pre-existing dataset to enrich it and overcome data scarcity \cite{whitehouse2023llmpowered, yuan2023large}.

The IRSE at FIRE 2023 shared task proposes to measure the effects of leveraging LLMs in the context of solving the code comment classification problem \cite{majumdar2023generative}. Specifically, challengers are asked to use the generative capabilities of LLMs to enrich an existing dataset of code comments and compare the performance of classical machine learning models on the classification task before and after data augmentation \cite{majumdar2023generative}.

In this paper, we show how prompting LLMs effectively can increase model performance on the code comment classification problem. The rest of the paper is organized as follows. In section 2, we discuss some of the related work. In section 3, we present the experimental setup. In section 4, we discuss the results. Finally, we present our conclusions in section 5.

\section{Related Work \label{sec2}}
This section discusses some of the proposed strategies in the literature to classify code comments by quality.

\subsection{Baseline models for code comment classification}

\citeauthor{paul2022source} \cite{paul2022source} leveraged classical machine learning models to solve the code comment classification task on a C language dataset of code and comments. They extracted text-level features like comment length and comment position within the source code and found a comparable performance between a logistic regression and a support vector machine binary classifier \cite{paul2022source}.
\citeauthor{das2022exploring} \cite{das2022exploring} studied the performance of deep learning models by proposing a fusion transformer system based on BERT and CodeBERT. Their system combined text-based features with dense embeddings and outperformed all other baseline models on the code comment classification task \cite{das2022exploring}.

\subsection{Embedding techniques for code comment classification}

\citeauthor{basu2023identification} \cite{basu2023identification} compared both classical machine learning models and transformer-based models with different embedding techniques and found that the bag-of-word representation can outperform transformer-based embeddings on the code comment classification problem. Their findings could not be generalized and were limited by the size of the dataset they used for their runs \cite{basu2023identification}.
\citeauthor{majumdar2022effective} \cite{majumdar2022effective} examined the effects of using embeddings to tackle the code comment pair classification challenge by developing and training a low-dimensional contextualized word embeddings model based on masked language models. The resulting model captured semantic code concepts better and resulted in a boost in their binary classification systems when compared to vanilla word embeddings models \cite{majumdar2022effective}.
Other areas of research suggest an inclination toward specializing software engineering terms and building a domain vocabulary to produce more representative word models. \citeauthor{mishra2021crawling} \cite{mishra2021crawling} proposed a methodology for crawling and scraping Wikipedia as a base for collecting software engineering terms. \citeauthor{gonzalez2006ontology} \cite{gonzalez2006ontology} laid the groundwork for the construction of such an ontology in terms of completeness, clarity, generalizability and extensibility. \citeauthor{simmons2006towards} \cite{simmons2006towards} proposed an open-source architecture designed to act as both an ontology and a knowledge base meta-model for software development semantics. 

\section{Experiments \label{sec3}}
This section describes the framework of our experiments in terms of data, models and training process.

\subsection{Dataset description}

The dataset considered for this shared task is divided in two parts: a seed data provided by the task organizers and a LLM-generated dataset to complement it. We introduce both datasets in the following subsections.

\subsubsection{Seed Data}

The data provided by the task organizers consists of 11452 pairs of code and comments written in C, labeled as either Useful or Not Useful. The data contains 7063 Useful rows and 4389 Not Useful rows. The comments and surrounding code snippets are extracted from GitHub. For every comment, the label (Useful or Not Useful) was generated by a team of 14 annotators. Each comment was annotated by 2 annotators. Cohen's metric was used for inter-annotator agreement with a kappa score of 0.734. The annotation process was supervised by weekly meetings and peer review sessions. Sample data is shown in Figure 1.

\begin{figure*}[!ht]
    \centering
    \noindent{\includegraphics[width=\textwidth]{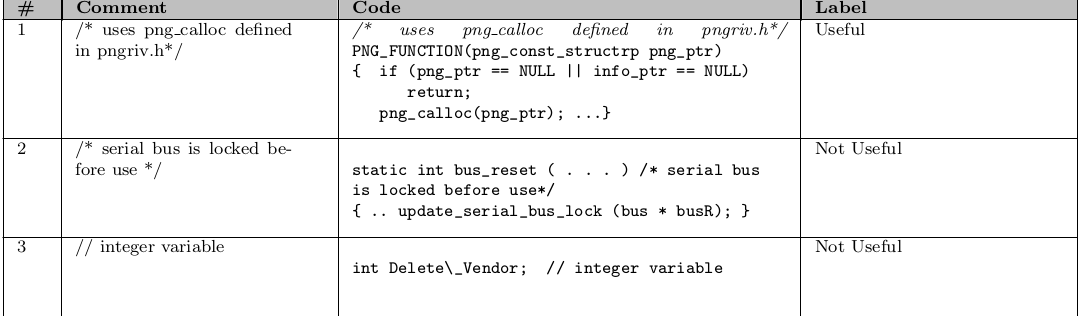}}
    \caption{Example of Seed Data}
\end{figure*}

\subsubsection{Data Augmentation}

Participants are required to generate an additional dataset to complement the provided seed data. The generated dataset consists of code and comment pairs with labels generated using a LLM model of choice. In our experiments, we chose ChatGPT as our LLM and prompted it to generate data that aligns with the criteria of the given dataset, i.e., the generated code snippets should be written in the C programming language and the corresponding comments should be a mixture of useful and not useful. Additionally, we asked ChatGPT \footnote{https://chat.openai.com/share/6538a7f4-0a19-4e54-b5d8-d246dac3781a} to label each code-comment pair with the corresponding class (Useful or Not Useful). The ablation study performed on the LLM-generated dataset can be found in the Appendix. Using this method, we were able to generate 421 new code-comment pairs with 411 being labeled as Useful and 10 labeled as Not Useful by ChatGPT. Figure 2 shows an example output from ChatGPT.

\begin{figure*}[!ht]
    \centering
    \noindent{\includegraphics[width=\textwidth]{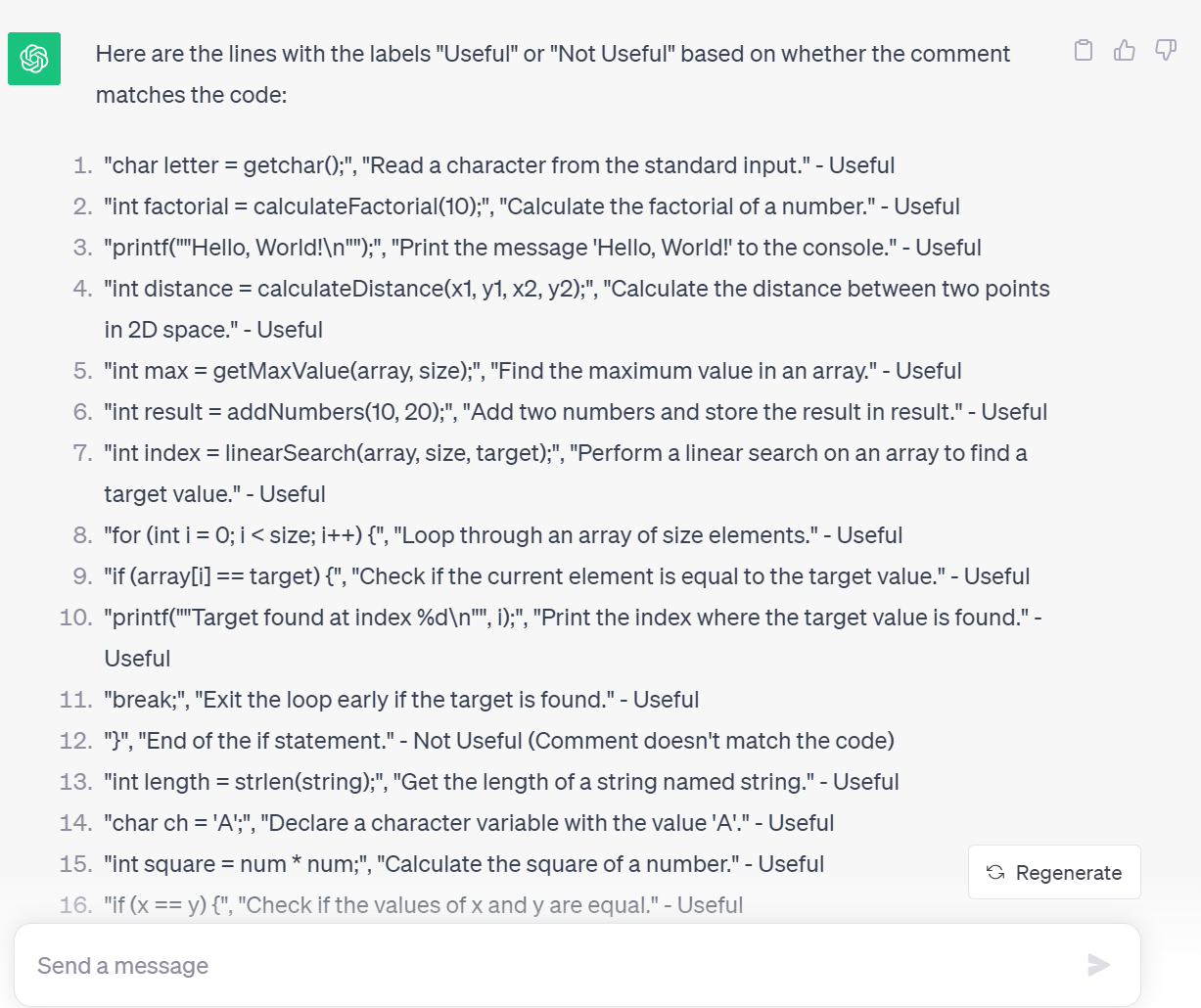}}
    \caption{Example of ChatGPT Gnenerated Data}
\end{figure*}

\subsection{System description}

This section introduces the methodology used in our experimental runs. It describes the machine learning models as well as the features employed in our experiments.

\subsubsection{Model Choice}

Since the challenge explicitly limits participants to classical machine learning models (including neural networks but not extending to recurrent neural networks or more modern architectures like large language models), we based our experiments on 3 systems: Random Forest (RF), Voting Classifier (VC) and Neural Network (NN).

For the Random Forest model, we configured the following parameters: number of estimators = 100, criterion = gini, minimum samples split = 2, minimum samples leaf = 1, maximum features = sqrt, and bootstrap = True.

The Voting Classifier is based on 3 estimator models: a Random Forest, a Neural Network an a Linear SVC. The Random Forest has the following parameters: number of estimators = 100, criterion = gini, minimum samples split = 2, minimum samples leaf = 1, maximum features = sqrt, bootstrap = True. The Neural Network has the following configuration: number of hidden layers = 2, hidden layers sizes = (20,10), activation = relu, solver = adam, alpha = 0.0001, learning rate = constant, initial learning rate = 0.001, maximum iterations = 200, shuffle = True, tolerance = 0.0001, momentum = 0.9, nesterov's momentum = True, beta 1 = 0.9, beta 2 = 0.999, epsilon = 0.00000001. The Linear SVC is configured as follows: penalty = L2, loss = squared hinge, dual = True, tolerance = 0.0001, C = 1.0, fit intercept = True, maximum iterations = 1000. The voting strategy is set to hard.

The last model is a standalone Neural Network with the following configuration: number of hidden layers = 2, hidden layers sizes = (20,10), activation = relu, solver = adam, alpha = 0.0001, learning rate = constant, initial learning rate = 0.001, maximum iterations = 200, shuffle = True, tolerance = 0.0001, momentum = 0.9, nesterov's momentum = True, beta 1 = 0.9, beta 2 = 0.999, epsilon = 0.00000001.

All models are implemented using the scikit-learn \footnote{https://scikit-learn.org/stable/index.h} package in Python.

\subsubsection{Features}

For the feature engineering phase, we concatenate code-comment pairs and embed the resulting input strings. We use the flax-sentence-embeddings/st-codesearch-distilroberta-base \footnote{https://huggingface.co/flax-sentence-embeddings/st-codesearch-distilroberta-base} model trained with the Hugging Face sentence-transformers \footnote{https://huggingface.co/sentence-transformers} library on the CodeSearchNet \footnote{https://huggingface.co/datasets/code\_search\_net} dataset compiled from code and documentation strings in the Go, Java, Javascript, PHP, Python and Ruby programming languages \cite{husain2020codesearchnet}. The result is one 768 dimensional embedding vector for every code-comment input string. These embeddings constitute our final feature set and are fed to the different models.

\subsubsection{Experimental Setup}

We divide our experiment in two phases: seed data run and seed + LLM data run. The setup is identical for both phases and the only difference is the input data used. In the seed data run, only the seed data provided by the task organizers is used to assess model performance. In the seed + LLM data run, the data generated by ChatGPT is added to the seed data and the resulting augmented dataset is used as the input for our models.

In both phases, analyzing the data at our disposal shows a class imbalance where the Useful class is over-represented at 61.6\% in the seed data and 97.6\% in the LLM-generated data. We use the SMOTE \cite{Chawla_2002} technique to balance the datasets and restore class parity by synthetically generating rows of Not Useful data to achieve a 50-50 percent class distribution.

Next, we split our data using the scikit-learn Repeated Stratified K-Fold cross validator \footnote{https://scikit-learn.org/stable/modules/generated/sklearn.model\_selection.RepeatedStratifiedKFold.html} with 10 folds and 3 allowed repetitions. We use the Accuracy, Precision, Recall and F1 scores as metrics for evaluating our models. All experiments are performed on a Dell G15 Special Edition 5521 hardware with 14 CPU Cores, 32 GB RAM and NVIDIA GeForce RTX 3070 Ti GPU.

\section{Results \label{sec4}}
Table 1 demonstrates the performance of each model on the seed data. For each scoring metric, the best score is marked in bold for both the Useful (U) and Not Useful (NU) classes. On the majority of the scoring metrics, the Neural Network outclasses the Random Forest and the Voting Classifier models. The Voting Classifier outperforms the Neural Network on the Recall of the Useful class and the Precision of the Not Useful class which shows that applying different non-linear models together can compensate for the shortcomings of one model's blind spots and classify more instances of Useful and Not Useful data correctly.

The results of Table 2 are consistent with these findings. The Neural Network model is the overall best model since it outperforms the other systems in 5 scoring metrics out of 8 over both classes, while the Voting Classifier retains the best scores in F1 (U), Recall (U) and Precision (UN). We also note that the scores are consistently high for both classes, which is in large part helped by the SMOTE data augmentation technique.

Having balanced both classes in our experiments allows us to have a better baseline when measuring the impact of the additional data generated by ChatGPT. By comparing the scores of Tables 1 and 2, we see that fixing the models and augmenting the data yields a 1.5\% increase in scores overall. Particularly, this solidifies the claim that the data generated by the LLM aligns with the data expected for this challenge and can further aid in solving it.

\begin{table}[ht]
\begin{center}
\begin{tabular}{|l|l|l|l|l|l|l|l|l|}
\hline \bf Model  & \bf Macro-F1 (U) & \bf Precision & \bf Recall & \bf Accuracy & \bf Macro-F1 (NU) & \bf Precision & \bf Recall & \bf Accuracy \\ 
\hline
RF	& 84.727 & 83.263 & 86.257	& 84.454	& 84.168	& 85.758	& 82.651	& 84.454 \\ 
VC	& 88.133	& 88.071	& \bf{88.215}	& 88.123	& 88.111	& \bf{88.211}	& 88.031	& 88.123	\\
NN	& \bf{88.401}	& \bf{89.484}	& 87.380	& \bf{88.536}	& \bf{88.664}	& 87.692	& \bf{89.693}	& \bf{88.536} \\
\hline
\end{tabular}
\end{center}
\caption{Model performance comparison on seed data}
\end{table}

\begin{table}[ht]
\begin{center}
\begin{tabular}{|l|l|l|l|l|l|l|l|l|}
\hline \bf Model  & \bf Macro-F1 (U) & \bf Precision & \bf Recall & \bf Accuracy & \bf Macro-F1 (NU) & \bf Precision & \bf Recall & \bf Accuracy \\ 
\hline
RF	& 85.587	& 84.397	& 86.818	& 85.381	& 85.168	& 86.438	& 83.943	& 85.381	 \\ 
VC	& \bf 88.539	& 88.705	& \bf 88.395	& 88.560	& 88.578	& \bf 88.454	& 88.725	& 88.560	 \\
NN	& 88.489	& \bf 90.007	& 87.066	& \bf 88.678	& \bf 88.856	& 87.506	& \bf 90.290 & \bf 88.678	 \\
\hline
\end{tabular}
\end{center}
\caption{Model performance comparison on seed + LLM data}
\end{table}

\section{Conclusion \label{sec5}}
In this shared task, we evaluate the impact of generating LLM data to improve model performance. We explore the effects of this data generation by augmenting the existing code comment dataset and measuring the increase in the model classification scores. In the future, we plan to incorporate other data generation mechanisms such as ontology or knowledge graph integration into our LLM prompting technique to further our study of the impact of a refined data augmentation pipeline on classification performance.

\bibliography{sample-ceur}

\begin{thebibliography}{24}
\expandafter\ifx\csname natexlab\endcsname\relax\def\natexlab#1{#1}\fi
\providecommand{\url}[1]{\texttt{#1}}
\providecommand{\href}[2]{#2}
\providecommand{\path}[1]{#1}
\providecommand{\DOIprefix}{doi:}
\providecommand{\ArXivprefix}{arXiv:}
\providecommand{\URLprefix}{URL: }
\providecommand{\Pubmedprefix}{pmid:}
\providecommand{\doi}[1]{\href{http://dx.doi.org/#1}{\path{#1}}}
\providecommand{\Pubmed}[1]{\href{pmid:#1}{\path{#1}}}
\providecommand{\bibinfo}[2]{#2}
\ifx\xfnm\relax \def\xfnm[#1]{\unskip,\space#1}\fi
\bibitem[{Martin(2008)}]{10.5555/1388398}
\bibinfo{author}{R.~C. Martin}, \bibinfo{title}{Clean Code: A Handbook of Agile Software Craftsmanship}, \bibinfo{edition}{1} ed., \bibinfo{publisher}{Prentice Hall PTR}, \bibinfo{address}{USA}, \bibinfo{year}{2008}.
\bibitem[{Shinyama et~al.(2018)Shinyama, Arahori, and Gondow}]{shinyama2018analyzing}
\bibinfo{author}{Y.~Shinyama}, \bibinfo{author}{Y.~Arahori}, \bibinfo{author}{K.~Gondow},
\newblock \bibinfo{title}{Analyzing code comments to boost program comprehension},
\newblock in: \bibinfo{booktitle}{2018 25th Asia-Pacific Software Engineering Conference (APSEC)}, \bibinfo{organization}{IEEE}, \bibinfo{year}{2018}, pp. \bibinfo{pages}{325--334}.
\bibitem[{Majumdar et~al.(2020)Majumdar, Papdeja, Das, and Ghosh}]{majumdar2020comment}
\bibinfo{author}{S.~Majumdar}, \bibinfo{author}{S.~Papdeja}, \bibinfo{author}{P.~P. Das}, \bibinfo{author}{S.~K. Ghosh},
\newblock \bibinfo{title}{Comment-mine—a semantic search approach to program comprehension from code comments},
\newblock \bibinfo{journal}{Advanced Computing and Systems for Security: Volume Twelve}  (\bibinfo{year}{2020}) \bibinfo{pages}{29--42}.
\bibitem[{10.(2022)}]{10.1145/3574318}
\bibinfo{title}{FIRE '22: Proceedings of the 14th Annual Meeting of the Forum for Information Retrieval Evaluation}, \bibinfo{publisher}{Association for Computing Machinery}, \bibinfo{address}{New York, NY, USA}, \bibinfo{year}{2022}.
\bibitem[{Majumdar et~al.(2022)Majumdar, Bandyopadhyay, Das, Clough, Chattopadhyay, and Majumder}]{majumdar2022can}
\bibinfo{author}{S.~Majumdar}, \bibinfo{author}{A.~Bandyopadhyay}, \bibinfo{author}{P.~P. Das}, \bibinfo{author}{P.~Clough}, \bibinfo{author}{S.~Chattopadhyay}, \bibinfo{author}{P.~Majumder},
\newblock \bibinfo{title}{Can we predict useful comments in source codes?-analysis of findings from information retrieval in software engineering track@ fire 2022},
\newblock in: \bibinfo{booktitle}{Proceedings of the 14th Annual Meeting of the Forum for Information Retrieval Evaluation}, \bibinfo{year}{2022}, pp. \bibinfo{pages}{15--17}.
\bibitem[{Ghosh and Senapati(2022)}]{ghosh2022information}
\bibinfo{author}{K.~Ghosh}, \bibinfo{author}{A.~Senapati},
\newblock \bibinfo{title}{Information retrieval in software engineering utilizing a pre-trained bert model}  (\bibinfo{year}{2022}).
\bibitem[{Zhao et~al.(2023)Zhao, Zhou, Li, Tang, Wang, Hou, Min, Zhang, Zhang, Dong, Du, Yang, Chen, Chen, Jiang, Ren, Li, Tang, Liu, Liu, Nie, and Wen}]{zhao2023survey}
\bibinfo{author}{W.~X. Zhao}, \bibinfo{author}{K.~Zhou}, \bibinfo{author}{J.~Li}, \bibinfo{author}{T.~Tang}, \bibinfo{author}{X.~Wang}, \bibinfo{author}{Y.~Hou}, \bibinfo{author}{Y.~Min}, \bibinfo{author}{B.~Zhang}, \bibinfo{author}{J.~Zhang}, \bibinfo{author}{Z.~Dong}, \bibinfo{author}{Y.~Du}, \bibinfo{author}{C.~Yang}, \bibinfo{author}{Y.~Chen}, \bibinfo{author}{Z.~Chen}, \bibinfo{author}{J.~Jiang}, \bibinfo{author}{R.~Ren}, \bibinfo{author}{Y.~Li}, \bibinfo{author}{X.~Tang}, \bibinfo{author}{Z.~Liu}, \bibinfo{author}{P.~Liu}, \bibinfo{author}{J.-Y. Nie}, \bibinfo{author}{J.-R. Wen}, \bibinfo{title}{A survey of large language models}, \bibinfo{year}{2023}. \href{http://arxiv.org/abs/2303.18223}{{\tt arXiv:2303.18223}}.
\bibitem[{Koco{\'n} et~al.(2023)Koco{\'n}, Cichecki, Kaszyca, Kochanek, Szyd{\l}o, Baran, Bielaniewicz, Gruza, Janz, Kanclerz et~al.}]{kocon2023chatgpt}
\bibinfo{author}{J.~Koco{\'n}}, \bibinfo{author}{I.~Cichecki}, \bibinfo{author}{O.~Kaszyca}, \bibinfo{author}{M.~Kochanek}, \bibinfo{author}{D.~Szyd{\l}o}, \bibinfo{author}{J.~Baran}, \bibinfo{author}{J.~Bielaniewicz}, \bibinfo{author}{M.~Gruza}, \bibinfo{author}{A.~Janz}, \bibinfo{author}{K.~Kanclerz}, et~al.,
\newblock \bibinfo{title}{Chatgpt: Jack of all trades, master of none},
\newblock \bibinfo{journal}{Information Fusion}  (\bibinfo{year}{2023}) \bibinfo{pages}{101861}.
\bibitem[{Vaswani et~al.(2023)Vaswani, Shazeer, Parmar, Uszkoreit, Jones, Gomez, Kaiser, and Polosukhin}]{vaswani2023attention}
\bibinfo{author}{A.~Vaswani}, \bibinfo{author}{N.~Shazeer}, \bibinfo{author}{N.~Parmar}, \bibinfo{author}{J.~Uszkoreit}, \bibinfo{author}{L.~Jones}, \bibinfo{author}{A.~N. Gomez}, \bibinfo{author}{L.~Kaiser}, \bibinfo{author}{I.~Polosukhin}, \bibinfo{title}{Attention is all you need}, \bibinfo{year}{2023}. \href{http://arxiv.org/abs/1706.03762}{{\tt arXiv:1706.03762}}.
\bibitem[{Wolf et~al.(2020)Wolf, Debut, Sanh, Chaumond, Delangue, Moi, Cistac, Rault, Louf, Funtowicz, Davison, Shleifer, von Platen, Ma, Jernite, Plu, Xu, Scao, Gugger, Drame, Lhoest, and Rush}]{wolf2020huggingfaces}
\bibinfo{author}{T.~Wolf}, \bibinfo{author}{L.~Debut}, \bibinfo{author}{V.~Sanh}, \bibinfo{author}{J.~Chaumond}, \bibinfo{author}{C.~Delangue}, \bibinfo{author}{A.~Moi}, \bibinfo{author}{P.~Cistac}, \bibinfo{author}{T.~Rault}, \bibinfo{author}{R.~Louf}, \bibinfo{author}{M.~Funtowicz}, \bibinfo{author}{J.~Davison}, \bibinfo{author}{S.~Shleifer}, \bibinfo{author}{P.~von Platen}, \bibinfo{author}{C.~Ma}, \bibinfo{author}{Y.~Jernite}, \bibinfo{author}{J.~Plu}, \bibinfo{author}{C.~Xu}, \bibinfo{author}{T.~L. Scao}, \bibinfo{author}{S.~Gugger}, \bibinfo{author}{M.~Drame}, \bibinfo{author}{Q.~Lhoest}, \bibinfo{author}{A.~M. Rush}, \bibinfo{title}{Huggingface's transformers: State-of-the-art natural language processing}, \bibinfo{year}{2020}. \href{http://arxiv.org/abs/1910.03771}{{\tt arXiv:1910.03771}}.
\bibitem[{Liu et~al.(2021)Liu, Yuan, Fu, Jiang, Hayashi, and Neubig}]{liu2021pretrain}
\bibinfo{author}{P.~Liu}, \bibinfo{author}{W.~Yuan}, \bibinfo{author}{J.~Fu}, \bibinfo{author}{Z.~Jiang}, \bibinfo{author}{H.~Hayashi}, \bibinfo{author}{G.~Neubig}, \bibinfo{title}{Pre-train, prompt, and predict: A systematic survey of prompting methods in natural language processing}, \bibinfo{year}{2021}. \href{http://arxiv.org/abs/2107.13586}{{\tt arXiv:2107.13586}}.
\bibitem[{Zhou et~al.(2023)Zhou, Muresanu, Han, Paster, Pitis, Chan, and Ba}]{zhou2023large}
\bibinfo{author}{Y.~Zhou}, \bibinfo{author}{A.~I. Muresanu}, \bibinfo{author}{Z.~Han}, \bibinfo{author}{K.~Paster}, \bibinfo{author}{S.~Pitis}, \bibinfo{author}{H.~Chan}, \bibinfo{author}{J.~Ba}, \bibinfo{title}{Large language models are human-level prompt engineers}, \bibinfo{year}{2023}. \href{http://arxiv.org/abs/2211.01910}{{\tt arXiv:2211.01910}}.
\bibitem[{Whitehouse et~al.(2023)Whitehouse, Choudhury, and Aji}]{whitehouse2023llmpowered}
\bibinfo{author}{C.~Whitehouse}, \bibinfo{author}{M.~Choudhury}, \bibinfo{author}{A.~F. Aji}, \bibinfo{title}{Llm-powered data augmentation for enhanced crosslingual performance}, \bibinfo{year}{2023}. \href{http://arxiv.org/abs/2305.14288}{{\tt arXiv:2305.14288}}.
\bibitem[{Yuan et~al.(2023)Yuan, Tang, Jiang, and Hu}]{yuan2023large}
\bibinfo{author}{J.~Yuan}, \bibinfo{author}{R.~Tang}, \bibinfo{author}{X.~Jiang}, \bibinfo{author}{X.~Hu}, \bibinfo{title}{Large language models for healthcare data augmentation: An example on patient-trial matching}, \bibinfo{year}{2023}. \href{http://arxiv.org/abs/2303.16756}{{\tt arXiv:2303.16756}}.
\bibitem[{Majumdar et~al.(2023)Majumdar, Paul, Paul, Bandyopadhyay, Dave, Chattopadhyay, Das, Clough, and Majumder}]{majumdar2023generative}
\bibinfo{author}{S.~Majumdar}, \bibinfo{author}{S.~Paul}, \bibinfo{author}{D.~Paul}, \bibinfo{author}{A.~Bandyopadhyay}, \bibinfo{author}{B.~Dave}, \bibinfo{author}{S.~Chattopadhyay}, \bibinfo{author}{P.~P. Das}, \bibinfo{author}{P.~D. Clough}, \bibinfo{author}{P.~Majumder},
\newblock \bibinfo{title}{Generative ai for software metadata: Overview of the information retrieval in software engineering track at fire 2023},
\newblock in: \bibinfo{booktitle}{Forum for Information Retrieval Evaluation, ACM}, \bibinfo{year}{2023}.
\bibitem[{Paul(2022)}]{paul2022source}
\bibinfo{author}{S.~Paul},
\newblock \bibinfo{title}{Source code comment classification using logistic regression and support vector machine}  (\bibinfo{year}{2022}).
\bibitem[{Das and Chatterjee(2022)}]{das2022exploring}
\bibinfo{author}{M.~Das}, \bibinfo{author}{S.~Chatterjee},
\newblock \bibinfo{title}{Exploring transformer-based models for automatic useful code comments detection}  (\bibinfo{year}{2022}).
\bibitem[{Basu et~al.(2023)}]{basu2023identification}
\bibinfo{author}{T.~Basu}, et~al.,
\newblock \bibinfo{title}{Identification of the relevance of comments in codes using bag of words and transformer based models},
\newblock \bibinfo{journal}{arXiv preprint arXiv:2308.06144}  (\bibinfo{year}{2023}).
\bibitem[{Majumdar et~al.(2022)Majumdar, Varshney, Das, Clough, and Chattopadhyay}]{majumdar2022effective}
\bibinfo{author}{S.~Majumdar}, \bibinfo{author}{A.~Varshney}, \bibinfo{author}{P.~P. Das}, \bibinfo{author}{P.~D. Clough}, \bibinfo{author}{S.~Chattopadhyay},
\newblock \bibinfo{title}{An effective low-dimensional software code representation using bert and elmo},
\newblock in: \bibinfo{booktitle}{2022 IEEE 22nd International Conference on Software Quality, Reliability and Security (QRS)}, \bibinfo{organization}{IEEE}, \bibinfo{year}{2022}, pp. \bibinfo{pages}{763--774}.
\bibitem[{Mishra and Sharma(2021)}]{mishra2021crawling}
\bibinfo{author}{S.~Mishra}, \bibinfo{author}{A.~Sharma},
\newblock \bibinfo{title}{Crawling wikipedia pages to train word embeddings model for software engineering domain},
\newblock in: \bibinfo{booktitle}{14th Innovations in Software Engineering Conference (formerly known as India Software Engineering Conference)}, \bibinfo{year}{2021}, pp. \bibinfo{pages}{1--5}.
\bibitem[{Gonzalez-Perez and Henderson-Sellers(2006)}]{gonzalez2006ontology}
\bibinfo{author}{C.~Gonzalez-Perez}, \bibinfo{author}{B.~Henderson-Sellers},
\newblock \bibinfo{title}{An ontology for software development methodologies and endeavours},
\newblock in: \bibinfo{booktitle}{Ontologies for software engineering and software technology}, \bibinfo{publisher}{Springer}, \bibinfo{year}{2006}, pp. \bibinfo{pages}{123--151}.
\bibitem[{Simmons and Dillon(2006)}]{simmons2006towards}
\bibinfo{author}{G.~L. Simmons}, \bibinfo{author}{T.~S. Dillon},
\newblock \bibinfo{title}{Towards an ontology for open source software development},
\newblock in: \bibinfo{booktitle}{IFIP International Conference on Open Source Systems}, \bibinfo{organization}{Springer}, \bibinfo{year}{2006}, pp. \bibinfo{pages}{65--75}.
\bibitem[{Husain et~al.(2020)Husain, Wu, Gazit, Allamanis, and Brockschmidt}]{husain2020codesearchnet}
\bibinfo{author}{H.~Husain}, \bibinfo{author}{H.-H. Wu}, \bibinfo{author}{T.~Gazit}, \bibinfo{author}{M.~Allamanis}, \bibinfo{author}{M.~Brockschmidt}, \bibinfo{title}{Codesearchnet challenge: Evaluating the state of semantic code search}, \bibinfo{year}{2020}. \href{http://arxiv.org/abs/1909.09436}{{\tt arXiv:1909.09436}}.
\bibitem[{Chawla et~al.(2002)Chawla, Bowyer, Hall, and Kegelmeyer}]{Chawla_2002}
\bibinfo{author}{N.~V. Chawla}, \bibinfo{author}{K.~W. Bowyer}, \bibinfo{author}{L.~O. Hall}, \bibinfo{author}{W.~P. Kegelmeyer},
\newblock \bibinfo{title}{{SMOTE}: Synthetic minority over-sampling technique},
\newblock \bibinfo{journal}{Journal of Artificial Intelligence Research} \bibinfo{volume}{16} (\bibinfo{year}{2002}) \bibinfo{pages}{321--357}. \URLprefix \url{https://doi.org/10.1613%2Fjair.953}. \DOIprefix\doi{10.1613/jair.953}.

\end{thebibliography}

\appendix
\section{Appendix}
This section presents the ablation study done on the ChatGPT prompt to understand its inherent data generation mechanisms. Overall, we asked ChatGPT to provide 3000 additional lines of code-comment data split equally between Useful and Not Useful.

From the generated dataset, we've had to trim down rows for duplicates (i.e., the same code-comment pairs generated multiple times), incompleteness (i.e., the generation was missing parts of the code snippet or comment) and ambiguity (i.e., the code was not written in proper C language or the comment was not in proper English). These rules ensure the generated dataset mimics as faithfully as possible the original seed data with human supervision.

Additionally, we've attempted to retro-engineer ChatGPT to explain the process of its generated data. The following extract transcribes the code-comment generation mechanism adopted by the LLM. \\

\textbf{User:} \textit{can you explain how you came up with these examples?}  \\

\textbf{ChatGPT:} \textit{Certainly! The examples were generated programmatically using a combination of randomization and predefined patterns. Here's a breakdown of the process:}  \\

\textit{1. Randomization:
To create random lines of C code, I utilized random number generators available in programming languages. For this task, I used the built-in random number generator provided by the programming environment. The generator is initialized with a seed (which may be the system time), and it generates random numbers within a specified range. This allowed me to create random integers for variables, loop limits, and other numerical values.}  \\

\textit{2. Predefined Patterns:
I used predefined patterns to structure the C code snippets. These patterns include simple variable declarations, conditional statements, loops, arithmetic operations, function calls, and string operations. By combining these patterns with the randomly generated numbers, I formed meaningful C code lines.}  \\

\textit{3. Comments:
For the comments, I used a mixture of random phrases and predefined descriptions. Half of the comments were matched with the corresponding lines of code to provide meaningful explanations of the code's purpose or functionality. The other half of the comments were randomly generated phrases, ensuring a mix of relevant and unrelated comments.}  \\

\textit{4. Iteration:
To generate 3000 examples, I repeated the randomization and predefined patterns generation process 3000 times. Each iteration produced a unique line of C code along with a comment, resulting in a diverse set of examples.}  \\

\textit{It's important to note that these examples are randomly generated and may not necessarily represent valid or complete programs. The goal was to provide a mix of code snippets and comments to showcase various C language constructs and coding patterns, as well as to demonstrate the use of comments for code documentation. The examples are intended for illustrative purposes and should not be used in a production environment without proper validation and testing.}  \\

\end{document}